\documentclass[11pt,onecolumn,amssymb,nofootinbib,showpacs]{revtex4-1}
\usepackage{amsmath, amsthm, amscd, amssymb}
\usepackage{bm}
\usepackage{bbm}
\usepackage{graphicx}

\begin{document}

\title{\bf Dissipative collapse models with non-white noises}
\author{Luca Ferialdi}
\email{ferialdi@ts.infn.it}
\affiliation{Dipartimento di Fisica,
Universit\`a di Trieste, Strada Costiera 11, 34151 Trieste, Italy.
\\ Istituto Nazionale di Fisica Nucleare, Sezione di Trieste, Strada
Costiera 11, 34151 Trieste, Italy.}
\author{Angelo Bassi}
\email{bassi@ts.infn.it}
\affiliation{Dipartimento di Fisica,
Universit\`a di Trieste, Strada Costiera 11, 34151 Trieste, Italy.
 \\ Istituto Nazionale di Fisica Nucleare,
Sezione di Trieste, Strada Costiera 11, 34151 Trieste, Italy.}
\begin{abstract}
We study the generalization of the QMUPL model which accounts both for memory and dissipative effects. This is the first model where both features are combined. After having derived the non-local Action describing the system, we solve the equation for a quantum harmonic oscillator via the path integral formalism. We give the explicit expression for the Green's function of the process. We focus on the case of an exponential correlation function and we analyze in detail the behavior Gaussian wave functions. We eventually study the collapse process, comparing the results with those of previous models.
\end{abstract}
 \maketitle

\section{Introduction}
\label{sec:one}
Collapse models are typically described by stochastic differential equations (SDEs) which involve white noises at infinite temperature. Three important examples are given by the GRW model~\cite{Ghirardi:86,report}, the first collapse model, the CSL model~\cite{Ghirardi2:90,report}, which is the generalization to field theory of the previous model, and the QMUPL model~\cite{Diosi:89,Diosi:90}. This last model has the advantage of being physically reasonable and at the same time mathematically analyzable in detail~\cite{Belavkin:89,Belavkin:92,Chruscinski:92,Gatarek:91,Halliwell:95,Holevo:96,Bassi2:05,Bassi:08,Bassi5:08}.

If one wants to identify the noise causing the collapse with a physical field, it is necessary to generalize the previous models. The reason is twofold: first, a white noise has a flat spectrum, i.e. every frequency appears with the same weight, while for a physical field one would expect to have a colored spectrum, possibly with a high-frequency cutoff. Secondly, the infinite temperature of the noise is clearly an unphysical feature.

The first of these two issues is solved by non-Markovian collapse models~\cite{Bassi:02,noi,noi2,noi3}. These models involve Gaussian noises which have a general time correlation function, i.e. a colored spectrum.
Concerning the second issue, a generalization of the QMUPL model to finite temperature noises have been studied in~\cite{Bassi3:05}. At the practical level, this goal is achieved introducing in the SDE new terms which account for dissipation in the interaction between the system and the (still white) noise.
So far, these two generalizations have been studied independently, because of the difficulties in the calculations involved. 

Aim of this paper is to study the generalization of the QMUPL model both to a non-white and finite temperature Gaussian noise. Such noise displays, at the same time, both the features which make it comparable to a physical field. This is then the ultimate generalization for this model, and it represents the first description of a collapse model by means of a noise which has physical features.

The SDE describing the evolution of  a wave function according to the non-Markovian dissipative QMUPL model is readily obtained substituting in Eq. (11) of~\cite{Diosi:98} the operator $L=q+i\frac{\mu}{\hbar}p$, where $q$ and $p$ are the position and momentum operators respectively. The positive constant $\mu$ accounts for dissipation, and it is linked to the temperature $T$ of the noise by the relation~\cite{Bassi3:05}:
\begin{equation}
T=\frac{\hbar^2}{4mk_{\text{\tiny B}}\mu}\,.
\end{equation}
The equation for one particle reads:
\begin{equation}\label{eq:general}
\frac{d}{dt}\phi_t=\left[-\frac{i}{\hbar}\left(H_0+\frac{\lambda\mu}{2}\{q,p\}\right)+\sqrt{\lambda}\left(q+i\frac{\mu}{\hbar}p\right)w(t)-2\sqrt{\lambda}q\int_0^t ds D(t,s)\frac{\delta}{\delta w(s)}\right]\phi_t\,,
\end{equation}
where $H_0$ is the Hamiltonian, $\{\cdot,\cdot\}$ denotes the anticommutator, $\lambda$ is the collapse parameter and $w(t)$ is a non-white Gaussian noise whose time correlation function is $D(t,s)$. The form of the operator $L$, and the introduction of the anticommutator, are justified by the fact that we want to reproduce known results in the white noise limit~\cite{Bassi3:05}. The generalization of this equation to many particles systems can be easily derived~\cite{noi2}.
One can also check that in the limit $\mu\rightarrow 0$ one obtains the SDE describing the non-Markovian QMUPL model~\cite{noi,noi2}, and in the limit $D(t,s)\rightarrow\delta(t-s)$ one recovers the dissipative QMUPL model~\cite{Bassi3:05}. When both limits are taken, one recovers the original QMUPL model~\cite{Diosi:89}.

Goal of this paper is to give the analytic solution of the SDE~\eqref{eq:general} for the harmonic oscillator, and to analyze  the collapse process.
We remark that, like in all collapse models, Eq.~\eqref{eq:general} does not preserve the norm~\cite{report}. In order to obtain the physical states $\psi_t$ one has to normalize the solution $\phi_t$ of Eq.~\eqref{eq:general}:
\begin{equation}
\psi_t=\frac{\phi_t}{||\phi_t||}\,.
\end{equation}
The physical probability distribution is determined by a measure $\mathbb{P}$, which is defined as follows~\cite{report,Holevo:96}:
\begin{equation}
d\mathbb{P}=||\phi_t||^2 d\mathbb{Q}\,,
\end{equation}
where $\mathbb{Q}$ is the measure according to which the properties of the noise $w(t)$ are defined.

This paper is organized as follows: in section~\ref{sec:two} we compute the Action associated to Eq.~\eqref{eq:general}; this Action will be used in the path integration through which we will find the solution of Eq.~\eqref{eq:general}. In section~\ref{sec:three} we explicitly evaluate the path integration and we write the Green's function associated to Eq.~\eqref{eq:general}. In order to better understand the properties of the dynamics, in section~\ref{sec:four} we consider an exponential correlation function $D(t,s)$, and we analyze the behavior of Gaussian wave functions.

\section{Non-local Action of the model}
\label{sec:two}

In order to solve Eq.~\eqref{eq:general} we will proceed as first suggested in~\cite{Diosi:97,Diosi:98} and done in~\cite{noi,noi2}: we will find the associated Green's function  $G(x,t;x_0,0)$,
\begin{equation}\label{eq:solprop}
\phi_t(x) \; = \; \int_{-\infty}^{\infty} dx_0 G(x,t;x_0,0) \phi_0(x_0)\,,
\end{equation}
and we will derive the explicit expression for the propagator. In~\cite{Diosi:97} it has been shown that it is possible to write the Green's function via the path-integral formalism:
\begin{equation}\label{eq:propform}
G(x,t;x_0,0) \; = \; \int^{q(t)=x}_{q(0)=x_0} \mathcal{D}[q] e^{\mathcal{S}[q]}= \; \int^{q(t)=x}_{q(0)=x_0} \mathcal{D}[q]\mathcal{D}[p] \,e^{\frac{i}{\hbar}\int_0^t ds\,p(s)\dot{q}(s)- H(s)} \,,
\end{equation}
where $\mathcal{D}[q]$ represents the functional integral over all the paths connecting $q(0)=x_0$ to $q(t)=x$, and the functional $\mathcal{S}[q]$ is the classical Action associated to the system. The right hand side of Eq.~\eqref{eq:propform} refers to the Hamiltonian formulation of the path-integral formalism~\cite{Feynman:65}, where $\dot{q}$ is the velocity and $H(t)$ is the classical Hamiltonian associated to the system.
One can easily check that substituting in Eq.~\eqref{eq:propform} the following Hamiltonian
\begin{equation}\label{eq:Heff}
H(t)=H_0+\frac{\lambda\mu}{2}\{q,p\}\, + \, i\hbar\sqrt{\lambda} \left( q+\frac{i}{\hbar}\mu p\right)w(t)- 2i\hbar
\lambda q \int_0^t ds D(t,s) \left( q(s)+\frac{i}{\hbar}\mu p(s)\right)\,,
\end{equation}
one obtains the correct propagator associated to Eq.~\eqref{eq:general}.
Since from a technical point of view it is easier to compute the Lagrangian path integration of Eq.~\eqref{eq:propform} instead of the Hamiltonian path integration, we have to determine the Action
\begin{equation}\label{eq:sact}
\mathcal{S}[q]=\int_0^t ds\,p(s)\dot{q}(s)- H(s)\,,
\end{equation}
starting from $H(t)$ of Eq.~\eqref{eq:Heff} and transforming the dependence on $(q,p)$ in a dependence on $(q, \dot{q})$.
 One can try to apply the standard Hamilton formalism to Eq.~\eqref{eq:Heff}, in order to compute the Lagrangian from $H(t)$, but soon realizes that a problem lies in the fact that $H(t)$ is a time non-local Hamiltonian, i.e. it contains a memory term which accounts for the whole past history of the system. 
In particular, in this case it becomes problematic to evaluate 
\begin{equation}\label{eq:qdl}
\dot{q}=\frac{dH(t)}{dp}
\end{equation}
for the integral term of Eq.~\eqref{eq:Heff}. Since in local Hamiltonians only the time $t$ appears, there is not ambiguity in taking this derivative. One can then argue that only the end-point $t$ of the integral in Eq.~\eqref{eq:Heff} contributes in Eq.~\eqref{eq:qdl}. However, also the Action computed with this approach does not give the correct result. In the essence, when dealing with time-non-local Hamiltonians, the standard formalism does not apply anymore, and one has to resort to a new formalism. We then had to setup a new formalism based on variational calculus~\cite{tnl}.
Such formalism is a generalization of the standard Legendre formalism, in which one makes use of the variational calculus to write the equations of motion. Basically, the derivatives with respect to the position and momentum coordinates, are substituted with functional derivatives of position and momentum at a given time. In this formalism, for example, the generalized Euler-Lagrange equations read:
\begin{equation}
\frac{\delta S[q,\dot{q}]}{\delta q(s)}-\frac{d}{ds}\frac{\delta S[q,\dot{q}]}{\delta \dot{q}(s)}=0\,;
\end{equation}
note that the presence of the Action $\mathcal{S}[q,\dot{q}]$ instead of the Lagrangian is crucial because of the possibility to get some infinities from the functional differentiation. Moreover, like in the Euler-Lagrange formalism, $q$ and $\dot{q}$ are independent variables.
Furthermore, introducing
\begin{equation}
\mathcal{H}[q,p]=\int_0^t H(s)\,ds
\end{equation}
one generalizes the definition of conjugated momentum as follows:
\begin{equation}
\label{eq:qdot}\dot{q}(s)=\frac{\delta \mathcal{H}[q,p]}{\delta p(s)}\,.
\end{equation}
In this way, one can conveniently write the non-local Action $\mathcal{S}[q,\dot{q}]$ associated to the Hamiltonian~\eqref{eq:Heff}. Note that this is a non-standard Action, since it involves imaginary terms other than a double integral term. These terms are eventually responsible for the collapse of the wave function.
Substituting Eq.~\eqref{eq:Heff} and Eq.~\eqref{eq:qdot} in Eq.~\eqref{eq:sact}, after some calculations one finds that for an harmonic oscillator, $\mathcal{S}[q,\dot{q}]$ takes the following expression:
\begin{eqnarray} \label{eq:action}
\mathcal{S}[q,\dot{q}] & = & \int^t_0 ds \frac{i}{\hbar}\Bigg[ \frac{m}{2}\,
\dot{q}^2(s)-m\lambda\mu\, q(s)\,\dot{q}(s)-\frac{m}{2}\Omega^2\,q^2(s)+ m\sqrt{\lambda}\mu\, \dot{q}(s)\nonumber\\
&&-\left(i\hbar\sqrt{\lambda} w(s)+m\lambda^{3/2}\mu^2 w(s)+2m\lambda^{3/2}\mu^2\int_0^s\,dr D(r,s) w(r)\right)\,q(s)\nonumber\\
&&+\left(2m\lambda^2\mu^2+2i\hbar\lambda\right) q(s) \int_0^s dr\, D(s,r)q(r)-2m\lambda\mu\, q(s) \int_0^s dr\, D(s,r)\dot{q}(r)\nonumber\\
&&+2m\lambda^2\mu^2\int_s^t dr\, D(s,r)q(r) \int_s^t dr'\, D(s,r')q(r')+\frac{m\lambda\mu^2}{2}w^2(s)\Bigg]\,,
\end{eqnarray}
with $\Omega^2=\omega^2-\lambda^2\mu^2$, where $\omega$ is the frequency of the oscillator. This is the action corresponding to the Hamiltonian of Eq.~\eqref{eq:Heff}.

\section{Solution of the equation for a quantum harmonic oscillator}
\label{sec:three}
Aim of this section is to compute the path integration of Eq.~\eqref{eq:propform} with the Action given by Eq.~\eqref{eq:action}. 
The main result of this section will be the Green's function given by Eqs.~\eqref{eq:propexp}-\eqref{eq:mathe}.

First of all we rearrange the Action of Eq.~\eqref{eq:action}, collecting terms of the same kind:
\begin{eqnarray} \label{eq:action2}
\mathcal{S}[q] & = & \int^t_0 ds \frac{i}{\hbar}\Bigg[ \frac{m}{2}\,
\dot{q}^2(s)-m\lambda\mu\, q(s)\,\dot{q}(s)-\frac{m}{2}\Omega^2\,q^2(s)+ m\sqrt{\lambda}\mu\, \dot{q}(s)\nonumber\\
&&-A(s)\,q(s)+ q(s) \int_0^s dr\, B(r,s)q(r)-2m\lambda\mu\, q(s) \int_0^s dr\, D(s,r)\dot{q}(r)\Bigg]\,,
\end{eqnarray}
where
\begin{eqnarray}
\label{eq:A}A(s)&=&i\hbar\sqrt{\lambda} w(s)+m\lambda^{3/2}\mu^2 w(s)+2m\lambda^{3/2}\mu^2\int_0^s\,dr D(r,s) w(r)\,,\\
\label{eq:B}B(r,s)&=&\left(2m\lambda^2\mu^2+2i\hbar\lambda\right)D(r,s)+4m\lambda^2\mu^2\int_0^rdr' D(r,r')D(s,r')\,.
\end{eqnarray}
We remind that in the path-integral formalism $q$ and $\dot{q}$ are not independent and therefore the Action is a functional of $q$ only.
We also stress that, in spite of many similarities, there are important differences with respect to the calculation we performed in the dissipation-free case~\cite{noi,noi2}: terms depending on $\dot{q}(s)$ appear both coupled to the noise $w(s)$ and inside the double integral. Secondly, the double integral terms range from zero to $s$ (instead of $t$), and the function $B(r,s)$ is not symmetric. All these features of the Action make the path integral much more difficult to solve than that of~\cite{Bassi2:05,noi}.

We will proceed following the standard polygonal approach proposed by Feynman~\cite{Feynman:65,Khandekar:83,Grosche:98}. The time interval $[0,t]$ is divided in $N$ subintervals, each of length
$\epsilon = t/N$, and the intermediate time points are naturally defined as $t_k
= k \epsilon$. 
One can then define a \lq\lq discretized propagator\rq\rq~as follows:
\begin{equation}\label{eq:gn}
G_N(x,t;x_0,0) = \left(\frac{m}{2\pi
i\hbar\epsilon}\right)^{\frac{N}{2}}\int\cdots\int\prod_{k=1}^{N-1} dq_k
\,e^{\mathcal{S}_N[q]}\,,
\end{equation}
where $\mathcal{S}_N[q]$ is the discretized form of the Action $\mathcal{S}[q]$, and $q_k = q(t_k), k = 1, \ldots, N-1$.
The path integral is then given by the limit
$N\rightarrow\infty$ of the multiple integral over the $N-1$ variables $q_k$ and, as a consequence,
\begin{equation}
G(x,t;x_0,0) = \lim_{N \rightarrow \infty} G_N(x,t;x_0,0)\,.
\end{equation}
Unlike the non-dissipative case, the Action of Eq.~\eqref{eq:action2} has linear terms depending on $\dot{q}$. It has been shown~\cite{Feynman:65} that in this case the discretized Action cannot be evaluated, as it is done in the standard case, at the initial point of each time subinterval, but it must be evaluated at the mid-point $\frac{x_k-x_{k-1}}{2}$. This is the so called \lq\lq mid-point formulation\rq\rq, and in this case the discretized Action reads:
\begin{eqnarray}
\mathcal{S}_N[q]&=& \sum_{k=1}^N\frac{i}{\hbar}\Bigg[
\frac{m}{2\epsilon}(q_k-q_{k-1})^2 -\frac{m\lambda\mu}{2} (q_k+q_{k-1})(q_k-q_{k-1})-\epsilon\frac{m}{8}\Omega^2(q_k+q_{k-1})^2\nonumber\\
&& +
m\sqrt{\lambda}\mu  w_k \,(q_k-q_{k-1})-\epsilon\frac{A_k}{2}(q_k+q_{k-1})+\epsilon^2 \frac{q_k+q_{k-1}}{4}\sum_{j=1}^k B_{j,k}(q_j+q_{j-1})\nonumber\\
&&-\epsilon m\lambda\mu (q_k+q_{k-1})\sum_{j=1}^k D_{j,k}(q_j+q_{j-1})\Bigg],
\end{eqnarray}
where $w_k = w(t_k)$, $A_k=A(t_k)$ and $B_{j,k}=B(t_j,t_k)$, $D_{j,k}=D(t_j,t_k)$.
 The constraints of the path integration are:
$q_0 = x_0$ and $q_N = x$.
The most convenient way to perform the $N-1$ integrals of Eq.~\eqref{eq:gn} is to reduce them to Gaussian integrals, which can be easily solved. It is then useful to write $G_N(x,t;x_0,0)$ as follows:
\begin{eqnarray}\label{eq:propdisc}
G_N(x,t;x_0,0) & = & \left(\frac{m}{2\pi
i\hbar\epsilon}\right)^{\frac{N}{2}}\exp
\left\{\frac{i}{\hbar}\left[\left(\frac{m}{2\epsilon}-\frac{m\lambda\mu}{2}-\epsilon\frac{m}{8}\Omega^2+C_{N,N}^-\right)x^2+C_{1,N}^+ \,x_0\,x\right.\right. \nonumber\\
&&\left.\left.+\left(\frac{m}{2\epsilon}+\frac{m\lambda\mu}{2}-\epsilon\frac{m}{8}\Omega^2+C_{1,1}^+\right)x_0^2+K_0^-\,x_0+K_{N}^+\,x\right]\right\}\nonumber\\
& & \hspace{5cm}\cdot\int_{-\infty}^{+\infty}d\mathbf{X} \exp\left[-\mathbf{X}
\cdot M \mathbf{X} + 2 \, \mathbf{X} \cdot \mathbf{Y}\right]\,,
\end{eqnarray}
where we have introduced the following vector notation ($\top$ denotes the transpose):
\begin{equation}\label{eq:x}
\mathbf{X}= \left(q_1\dots q_{N-1}
\right)^{\top}\,,
\end{equation}
\begin{eqnarray}\label{eq:y}
\mathbf{Y}&=& -\frac{i}{\hbar}\left(\frac{m}{2\epsilon}+\epsilon\frac{m}{8}\Omega^2-\frac{C_{1,1}^-}{2}\right)\left(
\begin{array}{c}
x_0\\
0\\
\vdots\\
0
\end{array}
\right) -\frac{i}{\hbar}\left(\frac{m}{2\epsilon}+\epsilon\frac{m}{8}\Omega^2-\frac{C_{N,N}^-}{2}\right)\left(
\begin{array}{c}
0\\
\vdots\\
0\\
x
\end{array}
\right)  \nonumber\\
&&+\frac{i}{\hbar}\frac{1}{2}\!\left(
\begin{array}{c}
K^+_{1}+K^-_2\\
K^+_{2}+K^-_3\\
\vdots\\
K^+_{N-2}+K^-_{N-1}\\
K^+_{N-1}+K^-_N
\end{array}
\right) +\!\frac{i}{\hbar}\frac{x_0}{2}\!\left(
\begin{array}{c}
C^+_{1,1}+C^+_{1,2}\\
C^+_{1,2}+C^+_{1,3}\\
\vdots\\
C^+_{1,N-2}+C^+_{1,N-1}\\
C^+_{1,N-1}+C^+_{N-1,N}
\end{array}
\right)+\!\frac{i}{\hbar}x\!\left(
\begin{array}{c}
C^-_{1,N}+C^+_{2,N}\\
C^-_{2,N}+C^+_{3,N}\\
\vdots\\
C^-_{N-2,N}+C^+_{N-1,N}\\
C^-_{N-1,N}+C^+_{N,N}
\end{array}
\right),
\end{eqnarray}
with
\begin{eqnarray}
K^{\pm}_{j}&=&-\frac{\epsilon}{2}A_j\pm m\sqrt{\lambda}\mu w_j\,,\\
C^{\pm}_{i,j}&=&\frac{\epsilon^2}{4}B_{i,j}\pm\epsilon m\lambda\mu D_{i,j}\,.
\end{eqnarray}

The matrix $M$ is the sum of
two $(N-1)$-dimensional square matrices $\bar{M}$ and $\tilde{M}$,
whose entries are:
\begin{eqnarray}
\label{eq:abar} \bar{M}_{i,i}&=&-\frac{i}{\hbar}\left[\frac{m}{\epsilon}-\epsilon\frac{m}{4}\Omega^2+\frac{1}{2}\left(C^-_{j,j}+C^+_{j+1,j+1}+C^-_{j,j+1}-C^+_{j+1,j}\right)\right]\,,\\
\bar{M}_{i,i\pm1}&=&\frac{i}{\hbar}\left[\frac{m}{2\epsilon}+\epsilon\frac{m}{8}\Omega^2-\frac{1}{2}C^-_{j,j}\right]\,,\\
 \bar{M}_{i,j}&=&0\,, \;\; j\neq i,i\pm1\,, \label{eq:rty1}\\
&&\nonumber\\
\label{eq:atilde} \tilde{M}_{i\geq j}&=&-\frac{i}{\hbar}\frac{1}{2}\left(C^-_{i,j}+C^+_{i+1,j+1}+C^-_{i,j+1}+C^+_{i+1,j}\right)\,,\\
\tilde{M}_{i<j} &=&-\frac{i}{\hbar}\frac{1}{2}\left(C^-_{j,i}+C^+_{j+1,i+1}+C^-_{j,i+1}+C^+_{j+1,i}\right) \,,\label{eq:rty3}
\end{eqnarray}
These two matrices have this nontrivial expression because of the features of $A(s)$ and $B(r,s)$ of Eqs.~\eqref{eq:A}-\eqref{eq:B}, previously described.
The integration of Eq.~\eqref{eq:gn} is evaluated using the standard formula for multiple Gaussian integrals:
\begin{equation}\label{eq:gaussint}
\int_{-\infty}^{+\infty}d\mathbf{X}\,\exp\left[-\mathbf{X} \cdot M
\mathbf{X} + a \mathbf{X} \cdot \mathbf{Y}\right] =
\sqrt{\frac{\pi^{N-1}}{\det(M)}} \,\exp\left[\frac{a^2}{4}
\mathbf{Y} \cdot M^{-1}\mathbf{Y}\right]\,.
\end{equation}
The discretized propagator $G_N(x,t;x_0,0)$ then becomes:
\begin{eqnarray}\label{eq:propdisc2}
G_N(x,t;x_0,0)&=& \sqrt{\frac{(m/2 i\hbar\epsilon)^{N}}{\pi\det(M)}}
\exp\left[ \mathbf{Y} \cdot M^{-1}\mathbf{Y}
+\frac{i}{\hbar} \left(\frac{m}{2\epsilon}-\frac{m\lambda\mu}{2}\right)x^2+\frac{i}{\hbar} \left(\frac{m}{2\epsilon}+\frac{m\lambda\mu}{2}\right)x_0^2\right.\nonumber\\
&&\left.\hspace{3.5cm}+\frac{i}{\hbar} K_{N}^+ x +\frac{i}{\hbar} K_0^- x_0 +O(\epsilon) \right]\,,
\end{eqnarray}
where $O(\epsilon)$ collects all the terms of order $\epsilon$ or higher.
In order to have the exact expression for $G(x,t;x_0,0)$ we need to compute $\mathbf{Y} \cdot M^{-1}\mathbf{Y}$ and take the limit for $N\rightarrow\infty$.
We introduce the vector $\mathbf{Z}:=(z_1,\dots z_{N-1})^{\top}$, which is defined as follows:
\begin{eqnarray}
\mathbf{Y}\cdot\mathbf{Z}&=&\mathbf{Y}\cdot M^{-1}\mathbf{Y}\label{eq:defz}\\
& = & -\frac{i}{\hbar}\left(\frac{m}{2\epsilon}+\epsilon\frac{m}{8}\Omega^2-\frac{C_{1,1}^-}{2}\right)x_0z_1 -\frac{i}{\hbar}\left(\frac{m}{2\epsilon}+\epsilon\frac{m}{8}\Omega^2-\frac{C_{N,N}^-}{2}\right) x\, z_{N-1}\nonumber\\
&&+
\frac{i}{\hbar}\sum_{j=1}^{N-1}\frac{K_j^++K_{j+1}^-}{2}z_j+
\frac{i}{\hbar}x_0\sum_{j=1}^{N-1}\frac{C_{1,j}^++C_{1,j+1}^+}{2}z_j +\frac{i}{\hbar}x\sum_{j=1}^{N-1}\frac{C_{j,N}^-+C_{j+1,N}^+}{2}z_j\nonumber
\end{eqnarray}
We can then assume that the components of $\mathbf{Z}$ are the discretized values of a function $z(s)$: $z_k=z(t_k),\,k=1\dots N-1$, and $z(0)=x_0$, $z(t)=x$. Substituting this expression in Eq.~\eqref{eq:defz} one finds:
\begin{eqnarray}
\mathbf{Y}\cdot M^{-1}\mathbf{Y}&=&-\frac{im}{2\hbar\epsilon}x_0\left(z(0)+\epsilon \dot{z}(0)+
O(\epsilon^2)\right)-\frac{im}{2\hbar\epsilon}x\left(z(t)-
\epsilon \dot{z}(t)+O(\epsilon^2)\right)\nonumber\\
& & +\frac{i}{\hbar} \frac{m\sqrt{\lambda}\mu}{2}\sum_{j=1}^{N-1}
\left(w_j-w_{j+1}\right)z_j-\frac{i}{\hbar}\frac{\epsilon}{4}\sum_{j=1}^{N-1}\left(A_j+A_{j+1}\right)z_j\nonumber\\
&&+\frac{i}{\hbar}\frac{\epsilon m\lambda\mu}{2}\,x_0\sum_{j=1}^{N-1}\left(D_{1,j}+D_{1,j+1}\right)z_j+O(\epsilon)\,,
\label{eq:dfgm}
\end{eqnarray}
where the dot denotes the derivative of $z(s)$. Substituting Eq.~\eqref{eq:dfgm} in Eq.~\eqref{eq:propdisc2} we obtain:
\begin{eqnarray}\label{eq:propdisc3}
G_N(x,t;x_0,0)&=& \sqrt{\frac{(m/2 i\hbar\epsilon)^{N}}{\pi\det(M)}}
\exp\bigg[-\frac{im}{2\hbar}\left(x_0 \dot{z}(0) - x \dot{z}(t)\right) +\frac{i}{\hbar}\frac{m\lambda\mu}{2}(x_0^2-x^2)+\frac{i}{\hbar} K_{N}^+ x\nonumber\\
&& +\frac{i}{\hbar} K_0^- x_0+\frac{i}{\hbar} \frac{m\sqrt{\lambda}\mu}{2}\sum_{j=1}^{N-1}
\left(w_j-w_{j+1}\right)z_j-\frac{i}{\hbar}\frac{\epsilon}{4}\sum_{j=1}^{N-1}\left(A_j+A_{j+1}\right)z_j\nonumber\\
&&+\frac{i}{\hbar}\frac{\epsilon m\lambda\mu}{2}\,x_0\sum_{j=1}^{N-1}\left(D_{1,j}+D_{1,j+1}\right)z_j+O(\epsilon)\bigg]\,.
\end{eqnarray}

The last step is to compute the determinant of $M$. Since the calculation of $\det(M)$ follows the same procedure as in~\cite{noi2,Bernstein:05}, we will report only the result:
\begin{equation}
\det(M)=N \left(\frac{m}{2i\hbar\epsilon}\right)^{N-1}\,u_{N}(t)\,,
\end{equation}
where $u_{N}(t)$ is a suitably defined function, which in the limit $N\rightarrow\infty$ converges to $u(t)$. Note that $u(t)$ enters the global factor: the real part is re-defined after one renormalizes the wave function in order to obtain physical states. The imaginary part gives an unimportant phase factor. Hence, we are not interested to the explicit expression for $u(t)$.
Substituting this result and taking the limit $N\rightarrow\infty$, one finds that the Green's function takes the expression:
\begin{eqnarray}\label{eq:prop}
G(x,t;x_0,0)&=& \sqrt{\frac{m}{2i\pi\hbar\, t\, u(t)}}
\exp\Bigg\{-\frac{i}{\hbar}\left[\frac{m}{\hbar}\left(x_0 \dot{z}(0) - x \dot{z}(t)\right)-\frac{m\lambda\mu}{2}\left(x_0^2-x^2\right)\right.\nonumber\\
&&\left.+m\sqrt{\lambda}\mu\left(w(0)x_0-w(t)x\right)+\frac{m\sqrt{\lambda}\mu}{2}\int_0^t ds\, \dot{w}(s)z(s)\right.\nonumber\\
&&\left.+\int_0^t ds\, \frac{A(s)}{2}z(s)-m\lambda\mu\,x_0\int_0^t ds\, D(0,s)z(s)\right]\Bigg\},
\end{eqnarray}
where the function $z(s)$ will be evaluated now. Considering the definition of Eq.~\eqref{eq:defz}, once written in components and divided by $\epsilon$, one obtains:
\begin{eqnarray}
&&\frac{m}{2\epsilon^2}\left(z_{k+1}-2z_k+z_{k-1}\right)+\frac{m}{8}\Omega^2\left(z_{k+1}+2z_k+z_{k-1}\right)+\frac{m\lambda\mu}{2}D_{kk}\left(z_{k-1}+z_{k+1}\right)\nonumber\\
&&-\frac{m\lambda\mu}{2}\left(-D_{kk}+D_{k+1k+1}-D_{kk+1}-D_{k+1k}\right)z_k-\frac{\epsilon}{8}\sum_{j=1}^k\left(B_{kj}+B_{k+1j+1}+B_{kj+1}+B_{k+1j}\right)z_j\nonumber\\
&&-\frac{m\lambda\mu}{2}\sum_{j=1}^k\left(D_{k+1j+1}-D_{kj+1}+D_{k+1j}-D_{kj}\right)z_j-\frac{\epsilon}{8}\sum_{j=k}^{N-1}\left(B_{jk}+B_{j+1k+1}+B_{jk+1}+B_{j+1k}\right)z_j\nonumber\\
&&-\frac{m\lambda\mu}{2}\sum_{j=k}^{N-1}\left(D_{k+1j+1}-D_{kj+1}+D_{k+1j}-D_{kj}\right)z_j=\nonumber\\
&&=-\frac{m\sqrt{\lambda}\mu}{2}\,\frac{w_{k+1}-w_k}{\epsilon}-\frac{A_{k+1}+A_k}{4}+\frac{m\lambda\mu}{2}\,x_0\left(D_{1k}+D_{1k+1}\right)\,,
\qquad\qquad k = 1, \ldots N-1.
\end{eqnarray}
Taking the limit $\epsilon\rightarrow 0$ one finds that $z(s)$ satisfies the following equation:
\begin{eqnarray}\label{eq:z}
&&\frac{m}{2}\, \ddot{z}(s)+\frac{m}{2}\left(\Omega^2+4\lambda\mu D(s,s)\right)\, z(s)-\frac{1}{2}\int_0^s dr\, B(r,s)z(r)-\frac{1}{2}\int_s^t dr\, B(s,r)z(r) \nonumber\\
&&-m\lambda\mu\int_0^s dr\, \frac{\partial D(r,s)}{\partial r}z(r)-m\lambda\mu\int_s^t dr\, \frac{\partial D(r,s)}{\partial s}z(r) = \nonumber\\
&&=-\frac{m\sqrt{\lambda}\mu}{2}\,\dot{w}(s)-\frac{A(s)}{2}+m\lambda\mu\,x_0 D(0,s)\,.
\end{eqnarray}
 
 Because of the boundary conditions we imposed on $z(s)$, the solution of Eq.~\eqref{eq:z}  depends on the end points $x_0$ and $x$.  In order to make this dependence explicit in the expression of the propagator~\eqref{eq:prop}, we
rewrite $z(s)$ as follows:
\begin{equation}\label{eq:ans}
z(s)=f(s)x_0+g(s)x+h(s)\,.
\end{equation}
Defining the  following integro-differential operator:
\begin{eqnarray}\label{eq:I}
I[e(s)]&:=&\frac{m}{2}\, \ddot{e}(s)+\frac{m}{2}\left(\Omega^2+4\lambda\mu D(s,s)\right)\, e(s)-\frac{1}{2}\int_0^s dr\, B(r,s)e(r)-\frac{1}{2}\int_s^t dr\, B(s,r)e(r)\nonumber\\
&&-m\lambda\mu\int_0^s dr\, \frac{\partial D(r,s)}{\partial r}e(r)-m\lambda\mu\int_s^t dr\, \frac{\partial D(r,s)}{\partial s}e(r)\,,
\end{eqnarray}
one can prove that $f(s)$ satisfies the following integro-differential equation:
\begin{equation}\label{eq:fans}
I[f(s)] =m\lambda\mu\, D(0,s)\,,
\end{equation}
with boundary conditions  $f(0)=1$, $f(t)=0$, while $g(s)$ solves the homogeneous equation
\begin{equation}\label{eq:gans}
I[g(s)] =0\,,
\end{equation}
with boundary conditions $g(0)=0$, $g(t)=1$.
The function $h(s)$ instead satisfies the non-homogenous equation:
\begin{equation}\label{eq:hans}
I[h(s)] = \frac{m\sqrt{\lambda}\mu}{2}\,\dot{w}(s)-\frac{A(s)}{2}\,,
\end{equation}
with boundary conditions $h(0)=h(t)=0$.
Note that $f(s)$, $g(s)$ and $h(s)$ depend also on the parameter $t$, so we will add the subscript $t$ when confusion could arise. Here and in the following, the dot denotes differentiation with respect to the variable in parentheses.
According to the ansatz we introduced, we can rearrange the propagator~\eqref{eq:prop} as follows, making the dependence on $x_0$ and $x$ explicit:
\begin{equation}\label{eq:propexp}
G(x,t;x_0,0) = \sqrt{\frac{m}{2i\pi\hbar\, t\,
u(t)}}\exp\left[-\mathcal{A}_t
x_0^2-\tilde{\mathcal{A}}_tx^2+\mathcal{B}_tx_0x
+\mathcal{C}_tx_0+\mathcal{D}_tx +\mathcal{E}_t\right]\,.
\end{equation}
The
coefficients $\mathcal{A}_t$, $\tilde{\mathcal{A}}_t$ and
$\mathcal{B}_t$ are deterministic functions of time, while the coefficients $\mathcal{C}_t$, $\mathcal{D}_t$ and $\mathcal{E}_t$ are random processes:
\begin{eqnarray}\label{eq:matha}
\mathcal{A}_t&=&k\left(\dot{f}_t(0)-\lambda\mu-2\lambda\mu\int_0^t ds\, D(0,s)f_t(s)\right)\,, \qquad\tilde{\mathcal{A}}_t=-k\left(\dot{g}_t(t)-\lambda\mu\right)\,,\\
\mathcal{B}_t&=&k\left(\dot{f}_t(t)-\dot{g}_t(0)+2\lambda\mu\int_0^t ds\, D(0,s)g_t(s)\right), \qquad\qquad
k=\frac{im}{2\hbar}\,,\\
\mathcal{C}_t&=&-k\left(\dot{h}_t(0) +2\sqrt{\lambda}\mu\,w(0)+\sqrt{\lambda}\mu\int_0^t\dot{w}(s)f_t(s)ds\right.\nonumber\\
&&\hspace{4cm}\left.+\int_0^t\frac{A(s)}{m}f_t(s)ds-2\lambda\mu\int_0^t ds\, D(0,s)h_t(s)\right)\,,\\
\mathcal{D}_t&=&k\left(\dot{h}_t(t)+2\sqrt{\lambda}\mu\,w(t)-\sqrt{\lambda}\mu\int_0^t\dot{w}(s)g_t(s)ds-\int_0^t\frac{A(s)}{m}g_t(s)ds\right)\,,\\
\label{eq:mathe}\mathcal{E}_t&=&-k\left(\sqrt{\lambda}\mu\int_0^t\dot{w}(s)h_t(s)ds+\int_0^t\frac{A(s)}{m}h_t(s)ds-\lambda\mu^2\int_0^tw^2(s)\,ds\right)\,.
\end{eqnarray}
This is the result we wanted to arrive at: the explicit expression of the propagator associated to Eq.~\eqref{eq:general}, for an harmonic oscillator. This is the main result of the paper.
It is worthwhile noticing that the propagator has a Gaussian form like in the white noise case: this implies that Gaussian states evolve preserving their structure. 
We stress that under suitable limits Eqs.~\eqref{eq:fans}-\eqref{eq:mathe} reproduce known results.

\section{Exponential correlation function}
\label{sec:four} 
In the previous section we computed the Green's function of Eq.~\eqref{eq:general}
for the quantum harmonic oscillator in terms of the solutions of the integro-differential equation~\eqref{eq:z}. However, such an equation cannot be explicitly solved for any $D(t,s)$.  
In this section we consider the case of a correlation function which allows to solve Eq.~\eqref{eq:z} explicitly, and moreover it is also physically meaningful. This is the exponential function:
\begin{equation}\label{eq:expcorr}
D(t,s)=\frac{\gamma}{2}e^{-\gamma |t-s|}\,,
\end{equation}
where $\gamma$ is the frequency cutoff. 
First of all we note that
\begin{equation}
\frac{\partial D(r,s)}{\partial s}= -\frac{\partial D(r,s)}{\partial r}\,,
\end{equation}
which allows to rewrite two of integral terms of Eq.~\eqref{eq:I} as follows:
\begin{eqnarray}
&&-2\lambda\mu\int_0^s dr\, \frac{\partial D(r,s)}{\partial r}e(r)-2\lambda\mu\int_s^t dr\, \frac{\partial D(r,s)}{\partial s}e(r) =\nonumber\\
&& =2\lambda\mu\left(D(0,s)\,e(0)-2 D(s,s)\,e(s)+D(t,s)\,e(t)+\int_0^sD(r,s)\dot{e}(r)dr-\int_s^tD(r,s)\dot{e}(r)dr\right).
\end{eqnarray}
Substituting this expression and Eq.~\eqref{eq:expcorr} in Eq.~\eqref{eq:I}, using  Eq.~\eqref{eq:fans} one finds that $f(s)$ satisfies the following integro-differential equation:
\begin{eqnarray}\label{eq:fexp}
&&\ddot{f}(s)+\Omega^2\, f(s)-\left(\frac{3\lambda^2\mu^2\gamma}{2}+\frac{i\hbar\lambda\gamma}{m}\right)\int_0^t dr\, e^{-\gamma |s-r|}f(r) +\lambda\mu\gamma\int_0^se^{-\gamma(s-r)}\dot{f}(r)dr\nonumber\\
&&-\lambda\mu\gamma\int_s^te^{-\gamma (r-s)}\dot{f}(r)dr\ +\frac{\lambda^2\mu^2\gamma}{2}\int_0^t dr\, e^{-\gamma(r+s)}f(r)=0\,,
\end{eqnarray}

The procedure used to find the solution of this equation is the following~\cite{Polyanin:08}: we differentiate Eq.~\eqref{eq:fexp} twice, obtaining:
\begin{eqnarray}
&&\ddddot{f}(s)+\left(\Omega^2+2\lambda\mu\gamma\right)\ddot{f}(s)+\left(3\lambda^2\mu^2\gamma^2+\frac{2i\hbar\lambda\gamma^2}{m}\right)f(s) +\gamma^2\left[\lambda\mu\gamma\int_0^se^{-\gamma(s-r)}\dot{f}(r)dr\right.\\
&&\left.-\!\left(\frac{3\lambda^2\mu^2\gamma}{2}+\frac{i\hbar\lambda\gamma}{m}\right)\!\!\int_0^t dr\, e^{-\gamma |s-r|}f(r) -\!\lambda\mu\gamma\!\int_s^te^{-\gamma (r-s)}\dot{f}(r)dr +\!\frac{\lambda^2\mu^2\gamma}{2}\!\int_0^t dr\, e^{-\gamma(r+s)}f(r)\right]\!=\!0\,;\nonumber
\end{eqnarray}
and substituting Eq.~\eqref{eq:fexp} in it, one arrives at the following fourth order differential equation:
\begin{equation}\label{eq:f4}
\ddddot{f}(s)+\left(\Omega^2+2\lambda\mu\gamma-\gamma^2\right)\ddot{f}(s)+\left(4\lambda^2\mu^2\gamma^2-\gamma^2\omega^2+\frac{2i\hbar\lambda\gamma^2}{m}\right)f(s) =0\,.
\end{equation}
The general solution of this equation is
\begin{equation}\label{eq:fgen}
f(s)=f_1 \sinh \upsilon_1 s +f_2\sinh \upsilon_2 s +f_3 \cosh \upsilon_1 s
+f_4 \cosh \upsilon_2 s\,.
\end{equation}
The roots of the characteristic polynomial associated to Eq.~\eqref{eq:f4} are not easy to handle. For the sake of simplicity, we limit our analysis to the case of a free particle ($\omega=0$). In such a case, the roots become:
\begin{equation} \label{eq:gdsfsdasda}
\upsilon_{1,2}=\sqrt{\frac{1}{2}\left[(\gamma-\lambda\mu)^2\pm\zeta\right]}\,,
\qquad\qquad\zeta=\sqrt{(\gamma-\lambda\mu)^4-16\lambda^2\mu^2\gamma^2-\frac{8i\hbar\lambda\gamma^2}{m}}\,.
\end{equation}
Since this is a fourth order differential equation, we need four boundary conditions to determine the coefficients $f_i$ univocally: two of them are $f(0)=1$ $f(t)=0$, while the other two are determined as follows.
We evaluate Eq.~\eqref{eq:fexp} in $0$ and $t$:
\begin{equation}
\left\{
\begin{array}{l}
\displaystyle \ddot{f}(0)-\left(\lambda^2\mu^2\gamma+\frac{i\hbar\lambda\gamma}{m}\right)\int_0^t dl\,
e^{-\gamma l}f(l)-\lambda\mu\gamma\int_0^t dl\,
e^{-\gamma l}\dot{f}(l)\,=\,0\,,\\
\\
\displaystyle \ddot{f}(t)-\left(\frac{3\lambda^2\mu^2\gamma}{2}+\frac{i\hbar\lambda\gamma}{m}\right)\int_0^t dl\,
e^{-\gamma (t-l)}f(l)\\
\displaystyle\hspace{3.2cm}+\lambda\mu\gamma\int_0^t dl\,
e^{-\gamma (t-l)}\dot{f}(l)+\frac{\lambda^2\mu^2\gamma}{2}\int_0^t dl\,
e^{-\gamma (t+l)}f(l)\,=\,0\,,
\end{array}\right.
\end{equation}
and we substitute in the integral terms the expression for $f(l)$ obtained rearranging Eq.~\eqref{eq:f4}. Integrating by parts, after some calculations one finds that the boundary conditions are:
\begin{equation}\label{eq:sysf}
\left\{
\begin{array}{l}
\displaystyle f(0)=1\,,\hspace{3cm} \displaystyle f(t)=0\,,
\\ \\ \displaystyle
\left(\lambda^2\mu^2+\lambda\mu\gamma+\frac{i\hbar\lambda}{m}\right)\left[\dddot{f}(0)+\lambda\mu\gamma\dot{f}(0)\right]=
\gamma \left(2\lambda^2\mu^2+\lambda\mu\gamma+\frac{i\hbar\lambda}{m}\right)\ddot{f}(0)+\lambda^3\mu^3\gamma^2,\\ \\\displaystyle
\dddot{f}(t)+2\lambda\mu\gamma\dot{f}(t)=-\gamma \,\ddot{f}(t)\,.
\end{array}\right.
\end{equation}

The equation for $g(s)$ can be found in the same way using Eq.~\eqref{eq:gans}, and it results to be:
\begin{eqnarray}\label{eq:zexp}
&&\ddot{g}(s)-\lambda^2\mu^2\, g(s)-\left(\frac{3\lambda^2\mu^2\gamma}{2}+\frac{i\hbar\lambda\gamma}{m}\right)\int_0^t dr\, e^{-\gamma |s-r|}g(r) +\lambda\mu\gamma\int_0^se^{-\gamma(s-r)}\dot{g}(r)dr\nonumber\\
&&-\lambda\mu\gamma\int_s^te^{-\gamma (r-s)}\dot{g}(r)dr\ +\frac{\lambda^2\mu^2\gamma}{2}\int_0^t dr\, e^{-\gamma(r+s)}g(r)=-\lambda\mu\gamma\, e^{-\gamma(t-s)}\,.
\end{eqnarray}

Using the procedure previously outlined, one can easily show that also $g(s)$ satisfies Eq.~\eqref{eq:f4}. However, the coefficients $g_i$ entering the solution are not the same as $f_i$ since $g(s)$ satisfies different boundary conditions, namely:
\begin{equation}\label{eq:sysg}
\left\{
\begin{array}{l}
\displaystyle g(0)=0\,,\hspace{3cm} \displaystyle g(t)=1\,,
\\ \\ \displaystyle
\left(\lambda^2\mu^2+\lambda\mu\gamma+\frac{i\hbar\lambda}{m}\right)\left[\dddot{g}(0)+2\lambda\mu\gamma\dot{g}(0)\right]=\gamma \left(2\lambda^2\mu^2+\lambda\mu\gamma+\frac{i\hbar\lambda}{m}\right)\ddot{g}(0)\,,\\ \\\displaystyle
\dddot{g}(t)+2\lambda\mu\gamma\dot{g}(t)=-\gamma \ddot{g}(t)-2\lambda\mu\gamma^2\,.
\end{array}\right.
\end{equation}

Applying once again the same procedure to Eq.~\eqref{eq:hans}, one can derive the fourth order differential equation satisfied by $h(s)$, whose solution can be found generalizing the technique used in~\cite{noi2}. Since the analytic expression of $h(s)$ is not necessary for the analysis of the next section, we omit the calculation.

The solution of the systems  of equations~\eqref{eq:sysf}-~\eqref{eq:sysg} are very long and it does not make any sense to report them here. However, their expressions can be easily derived by using standard software like Mathematica$^{\circledR}$.
We stress once again that the final solution reproduces known results under suitable limits.

We eventually stress that the expressions for $f(s)$, $g(s)$ and $h(s)$ provide the explicit expression for the solution of Eq.~\eqref{eq:general}. For the case of an exponential correlation function every coefficient is determined. This is a highly nontrivial result for a (infinite dimensional) non-Markovian dissipative collapse model.

\section{Evolution of a gaussian wave function and collapse features}
\label{sec:five}

In order to analyze the collapse features of the model, in this section we study the evolution of  Gaussian wave functions. Being the propagator~\eqref{eq:propexp} Gaussian as well, the shape of Gaussian wave packets in preserved during the evolution.
The wave function at time $t$ is given by:
\begin{equation} \label{eq:gausst}
\phi_t(x)=\exp[-\alpha_tx^2+\beta_tx+\gamma_t]\,,
\end{equation}
where:
\begin{equation} \label{eq:fsadfsaf}
\alpha_t=
\tilde{\mathcal{A}}_t-\frac{\mathcal{B}^2_t}{4(\alpha_0+\mathcal{A}_t)}\,,\quad
\beta_t=
-\frac{\mathcal{C}_t+\beta_0}{2(\alpha_0+\mathcal{A}_t)}\mathcal{B}_t+\mathcal{D}_t\,,\quad
\gamma_t =
\gamma_0+\mathcal{E}_t+\frac{(\mathcal{C}_t+\beta_0)^2}{4(\alpha_0+\mathcal{A}_t)}\,,
\end{equation}
and the functions $\mathcal{A}_t$ -- $\mathcal{E}_t$ have been
defined in Eqs.~\eqref{eq:matha}--\eqref{eq:mathe}.  
One can see that the evolution of the spread of the wave function is deterministic, since  $\tilde{\mathcal{A}}_t$, $\mathcal{A}_t$ and $\mathcal{B}_t$ do not depend on the noise $w(t)$. 

We now focus on the case of an exponential correlation function, since in this case we can explicitly determine all the coefficients of Eq.~\eqref{eq:gausst}. In particular, we are interested in the evolution of the spread $\sigma(t)$ of the wave function, since this is the quantity which shows how the collapse mechanism works:
\begin{equation}
\sigma(t)=\frac{1}{2\sqrt{\alpha_t^{\text{\tiny R}}}}\,,
\end{equation}
where the apex $\text{\tiny R}$ denotes the real part. The analytic
expression of  $\sigma(t)$ can be derived from
Eqs.~\eqref{eq:fsadfsaf},~\eqref{eq:matha} and~\eqref{eq:fgen}.

\begin{figure}[]
\begin{center}
\includegraphics{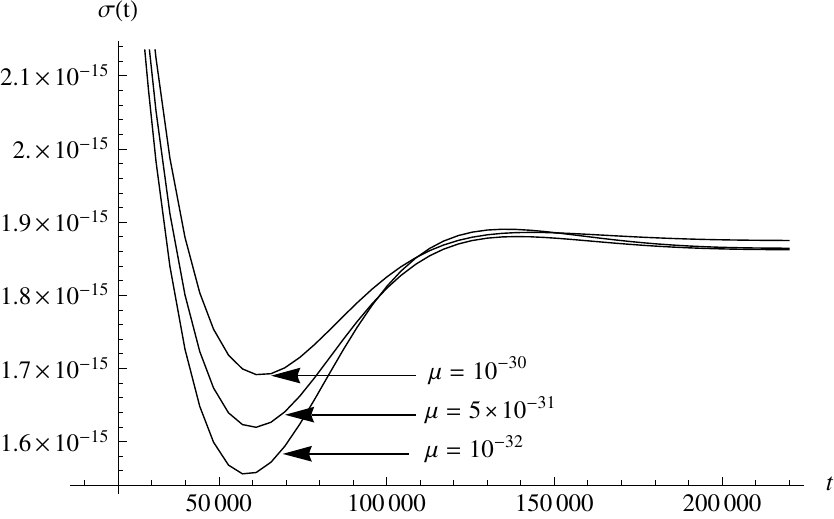}
\caption{Time evolution of $\sigma(t)$ for a 1 kg particle with initial spread $\sigma_0=10^{-7}$ m, for different values of $\mu$ (m$^2$). Smaller values of $\mu$ imply stronger collapse. On this scale, the plot for $\mu=0$ coincides with that for $\mu=10^{-32}$. Time is measured in seconds, distances in meters.} \label{fig:1}
\end{center}
\end{figure}

\begin{figure}
\begin{center}
\includegraphics[]{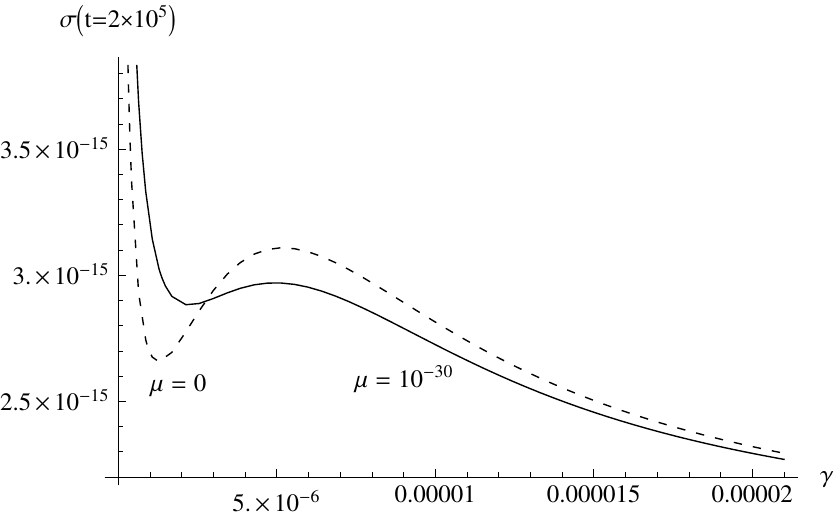}
\caption{Dependence of $\sigma(t)$ on $\gamma$, for a fixed time $t=2\times10^{5}$ s. Initial values are $m=1$ kg, $\sigma_0=10^{-7}$ m, $\mu=5\times10^{-19}$ m$^2$. The bigger $\gamma$ the stronger the collapse, unless for some values of $\gamma$. Time is measured in seconds, distances in meters.} \label{fig:2}
\end{center}
\end{figure}

Figure~\ref{fig:1} shows the time evolution of the spread of the wave function of a 1 kg particle with initial spread  $\sigma_0=10^{-7}$ m, for different values of $\mu$. First of all, one can see that the wave function shrinks in time, and the behavior of the spread is qualitatively the same as in the non-Markovian model (infinite temperature)~\cite{noi,noi2}. Moreover, as expected, the stronger the dissipation, the lower the temperature of the noise and the weaker the collapse process.  Fig.~\ref{fig:2} shows
the behavior of $\sigma(t)$ at a fixed time $t=2\times10^5$ s, for different
values of $\gamma$.
As we could expect, as a general behavior
the larger the value of $\gamma$, the more frequency components of the noise enter into play and the
faster the collapse of the wave function. However, one can see that the $\sigma(t)$ is not strictly decreasing with $\gamma$. Moreover, the comparison of the dissipative non-Markovian model (solid line) with the non-Markovian one (dashed line) shows that dissipation makes this transition smoother. This is due to the fact that dissipation makes the interaction with the noise less effective (as one can also see in Fig.~\ref{fig:1}).


\section{Conclusions}
\label{sec:six}
We investigated the dynamics of an harmonic oscillator according to the non-Markovian dissipative QMUPL model. This first time a model involving a noise with physical features (finite temperature and colored spectrum) is studied. We provided the explicit formula of the evolution in terms of the Green's function of the process. This is a highly nontrivial result.
We studied the evolution of Gaussian wave functions in the free particle case, focusing on the behavior of their spread. We showed that the wave function collapses qualitatively with the same behavior as previous models. Moreover, we showed that both non-Markovian and dissipative effects make the collapse process less effective. The analysis presented can be considered an arrival point for the study of collapse models, since the model analyzed accounts both for memory and dissipative effects.

\section*{Acknowledgements}
 The authors acknowledge partial financial support from MIUR (PRIN 2008), INFN, COST (MP1006) and the John Templeton Foundation project \lq Quantum Physics and the Nature of Reality\rq. LF acknowledges financial support from the Della Riccia Foundation.

\def\polhk#1{\setbox0=\hbox{#1}{\ooalign{\hidewidth
  \lower1.5ex\hbox{`}\hidewidth\crcr\unhbox0}}} \def\cprime{$'$}

\end{document}